\documentclass[final,3p,sort&compress]{elsarticle}
\usepackage{graphicx}
\usepackage[fleqn]{amsmath}

\def\be{\begin{equation}}
\def\ee{\end{equation}}
\def\ba{\begin{eqnarray}}
\def\ea{\end{eqnarray}}
\def\bd{\begin{displaymath}}
\def\ed{\end{displaymath}}
\def\bq{\begin{eqnarray}}
\def\eq{\end{eqnarray}}

\journal{Annals of Physics}

\begin{document}
\begin{frontmatter}

\title{Revisiting the quantum Szilard engine with fully quantum considerations}

\author[bit,sd]{Hai Li}
\author[bit]{Jian Zou \corref{corresp}}
\ead{zoujian@bit.edu.cn}
\author[bit]{Jun-Gang Li}
\author[bit]{Bin Shao}
\author[wla]{Lian-Ao Wu}
\cortext[corresp]{Corresponding author}

\address[bit]{School of Physics, Beijing Institute of Technology, Beijing 100081, China}
\address[sd]{School of Information and Electronics Engineering, Shandong Institute of
             Business and Technology, Yantai 264000, China}
\address[wla]{ Department of Theoretical Physics and History of Science,
The Basque Country University (EHU/UPV), P. O. Box 644, ES-48080
Bilbao, Spain and IKERBASQUE, Basque Foundation for Science,
ES-48011 Bilbao, Spain}

\date{\today}

\begin{abstract}
By considering level shifting during the insertion process we
revisit the quantum Szilard engine (QSZE) with fully quantum
consideration. We derive the general expressions of the heat
absorbed from thermal bath and the total work done to the
environment by the system in a cycle with two different cyclic
strategies. We find that only the quantum information contributes
to the absorbed heat, and the classical information acts like a
feedback controller and has no direct effect on the absorbed heat.
This is the first demonstration of the different effects of
quantum information and classical information for extracting heat
from the bath in the QSZE. Moreover, when the well width
$L\rightarrow \infty $ or the temperature of the bath
$T\rightarrow \infty $ the QSZE reduces to the classical Szilard
engine (CSZE), and the total work satisfies the relation
$W_{\mathtt{tot}}=k_{B}T \mathtt{ln}2$ as obtained by Sang Wook
Kim et al. [Phys. Rev. Lett. 106, 070401 (2011)] for one particle
case.

\end{abstract}

\begin{keyword}
 Quantum Szilard engine\sep
 Energy level shifts \sep
 Measurement\sep
 Quantum information
\end{keyword}

\end{frontmatter}

\section{Introduction}
\label{sec1}

Maxwell's Demon could separate hot atoms from cold, and therefore
could obtain work from a single heat bath. This seemed to violate
the second law of thermodynamics \cite{[1]Leff,[2]Maruyama} and
led to discussions and confusions until 1929, when Szilard devised
his \textquotedblright Szilard Engine\textquotedblright\ (SZE)
\cite{[3]Szilard}. The SZE could extract work from a bath using
classical information (acquired by measurement of the atom) and
establish the connection between work and entropy to reassure the
validation of the second law of thermodynamics, as illustrated in
Fig. 1. Later, Landauer \cite{[4]Landauer} and Bennett
\cite{[5]Bennett} completely analyzed the SZE, and showed that the
erasure or reset of the Demon memory costs at least the energy of
$k_{B}T\mathtt{ln}2$ associated with the entropy decrease of the
engine. It was conjectured \cite{[p]L. B.1,[p]L. B.2,[p]L. B.3}
that there exists a general equivalence relation between
information and work; namely, that by having any information $J$
about the state of a physical system, it is possible, by allowing
the system to relax to its maximum-entropy state, to convert into
mechanical work an amount of heat $W = k_BTJ$ without any entropy
increase in the environment. Moreover, there are many works on the
relationship between information and work, and some significant
results have been obtained \cite{[p]Scully,[p]Zureck,[p]Dahlsten}.

Research interests in the SZE have recently been revived in
various theoretical contexts
\cite{[6]Scully,[7]Kim,[8]Kim,[8]Raizen,[9]Marathe} and
experimental implementations
\cite{[9]Serreli,[10]Thorn,[11]Price}. However, early literatures
hardly paid attention to fully quantum analysis except for those
via measurement process \cite{[12]Zurek,[13]Lloyd}. Sang Wook Kim
noticed that work is required in the process of insertion for a
quantum Szilard engine (QSZE) \cite {[16]Kim}. This makes the
engine substantially different from its classical counterpart,
classical SZE (CSZE). It has been shown in Ref. \cite {[16]Kim}
that work in the insertion, expansion, removal processes, and the
entire
cycle are $W_{\mathtt{ins}}=-\Delta +k_{B}T\mathtt{ln}2$, $W_{\mathtt{exp}%
}=\Delta $, $W_{\mathtt{rem}}=0$ and
$W_{\mathtt{tot}}=k_{B}T\mathtt{ln}2$, respectively, when the
insertion process is performed isothermally.
 Here, $\Delta =\mathtt{ln}[\frac{z(L)}{Z(L/2)}]$, $%
z(l)=\sum_{n=1}^{\infty }e^{-\beta E_{n}(l)}$ and $E_{n}(l)=\frac{h^{2}n^{2}%
}{8ml^{2}}(n=1,2,3\dots )$ with $h$ and $m$ being Planck's
constant and the mass of the particle. In the QSZE, the insertion
of the wall is characterized by increase of the height of
potential barrier. Energy levels in the box vary with the boundary
conditions, contributing to the quantum thermodynamic work and the
system's internal energy. Both the position of the insertion and
the rate of increase of the potential barrier height influence the
level shifts. The faster the height of the potential barrier
increases, the greater the change of internal energy of the system
and the energy becomes infinite when the height tends to infinity
instantaneously, i.e., the insertion is carried out
instantaneously, as discussed in \cite{[14]Bender}. If the system
is initially in the ground state and the insertion is performed
adiabatically with the barrier being not at the center of the box,
the particle will end up definitely in the larger part of the box
\cite {[14]Bender} which is different from the classical
situation. In the case of isothermal insertion the effect of
energy level shifts is concealed by the heat exchange. So in order
to demonstrate the quantum effects of the QSZE completely, it is
necessary to consider the adiabatic insertion. In this paper we
assume that the insertion is performed adiabatically and analyze
the cycle of the QSZE with fully quantum considerations. It is
interesting to note that energy level shifts caused by the
boundary conditions during the insertion process play a
significant role in extracting work or absorbing heat in the QSZE.

\begin{figure}[tbp]
\centering \includegraphics[width=10.0cm]{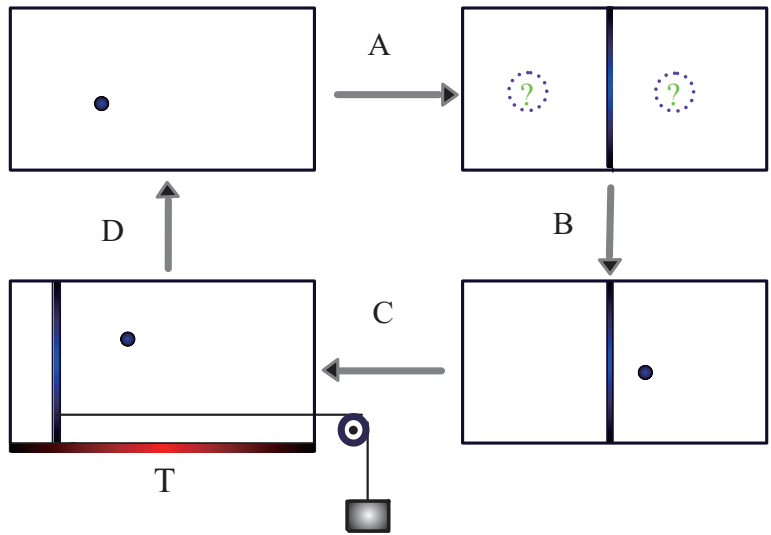} \hspace{0.5cm}
\caption{(Color online) Schematic diagram of the thermodynamic
processes of the CSZE. Initially a single particle is prepared in an
isolated box. (A) The box is divided into two equal subspaces by a
wall inserted at the center of box. The dotted circles indicate that
before the measurement which subspace the particle locates in is
unconfirmed. (B) The particle is found on one side after the
measurement. (C) A load is attached to the wall, and the particle
absorbs heat and does work via an isothermal expansion at a constant
temperature $T$. (D) To remove the wall which stops at the left end
of box, the box returns to the initial situation.}
\end{figure}

In this paper, we revisit the QSZE with fully quantum-mechanical
consideration. We consider a single particle in one-dimensional
infinite square well, and devise two different cyclic strategies.
For these two cyclic strategies we are able to know explicitly the
quantities of work done by the system, heat transferred, and the
change in internal energy in each step. In this way we can derive
the general expressions of heat transferred from the bath and the
total work done by the system. We find that the quantum
information plays a decisive role in the whole cycle and is
associated with heat absorbed from the bath and work done by the
system. However, the classical information of the particle being
located at seems to behave like a feedback controller, and has no
direct effect on heat absorption. This is the first demonstration
of the different effects between quantum information and classical
information for extracting heat from the bath in the QSZE. When
the well width $L\rightarrow \infty $ and bath temperature
$T\rightarrow \infty $, our results show that the QSZE reduces to
the CSZE.


The paper is organized as follows. We introduce our model of the
QSZE in section 2. We will present two different cyclic strategies
for the QSZE: One with isothermal expansion and the other with
adiabatic expansion, and analyze the cyclic processes of the QSZE
with fully quantum consideration in section 3. Two limits of the
QSZE at $ L\rightarrow \infty $ and $T\rightarrow \infty $ are
discussed in section 4. Finally, we present our conclusions in
section 5. Remarks on notational details and some technical
derivations are given in the appendixes.



\section{The model}\label{sec:mod}

Consider a single particle of mass $m$ confined to a one-dimensional
infinite square well of width $L$. The eigenvalues $E_{n}$ and eigenstates $%
|E_{n}\rangle $ are
\begin{equation}
E_{n}(L)=\frac{{n^{2}\hbar ^{2}\pi
^{2}}}{{2mL^{2}}},~~n=1,2,3...,
\label{1}
\end{equation}
\begin{equation}
\left|E_{n}(L)\right\rangle =\left
\{\begin{aligned}
\sqrt{\frac{2}{L}}\sin[\frac{{n\pi}(x-L/2)}{L}],~~~ &n=2k\\
\sqrt{\frac{2}{L}}\cos[\frac{{n\pi}(x-L/2)}{L}],~~~&n=2k-1
 \end{aligned}
 \right.,
\end{equation}%
where $k$ is a positive integer and $0\leq x\leq L$.

Assume that the system is initially in thermal equilibrium with a
bath at temperature $T$, the density matrix $\rho _{0}(L)$ reads as
\begin{equation}
\rho _{0}(L)=\sum_{n=1}^{\infty }P_{n}(L)|E_{n}(L)\rangle \langle
E_{n}(L)|,\label{5}
\end{equation}%
where $P_{n}(L)=\frac{{e^{-\beta E_{n}}}}{{Z(L)}}$ is the
probability of the particle in the eigenstate $|E_{n}\rangle $, and
satisfies the normalization condition $\sum_{n=1}^{\infty
}P_{n}(L)=1$. $Z(L)=\sum_{n=1}^{\infty }e^{-\beta E_{n}}$ is the
partition function, $\beta =\frac{1}{{k_{B}T}}$
and $k_{B}$ is the Boltzmann constant. The initial system's internal energy $%
U_{0}(L)$ and the initial Von-Neumann entropy $S_{0}$, are given by
\begin{equation}
U_{0}(L)=\sum_{n=1}^{\infty }P_{n}(L)E_{n}(L),
\label{6}
\end{equation}%
\begin{equation}
S_{0}=-k_{B}\mathtt{Tr}(\rho
_{0}\mathtt{ln}\rho_{0})=-k_{B}\sum_{n=1}^{\infty
}P_{n}(L)\mathtt{ln}P_{n}(L),
\label{7}
\end{equation}%
respectively.

\begin{figure}[tbp]
\centering \includegraphics[width=12cm]{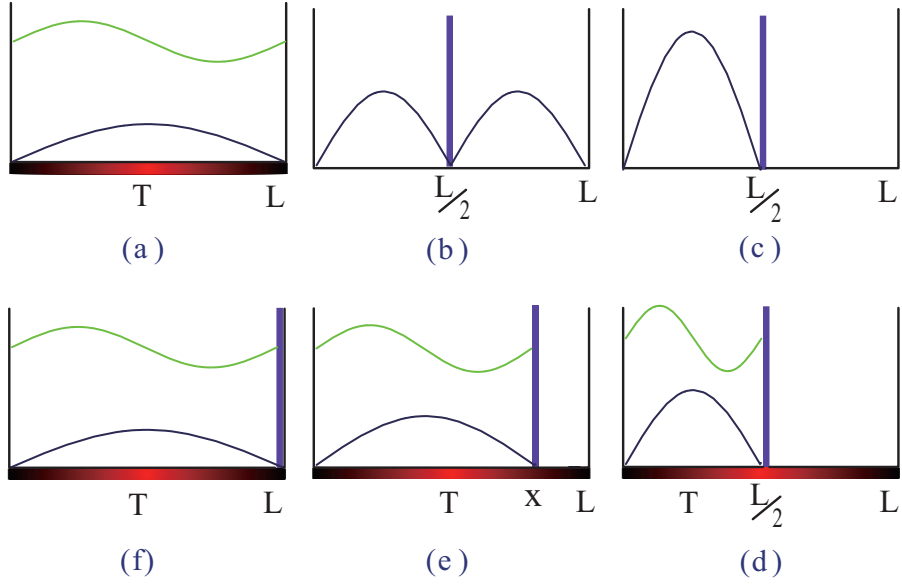} \hspace{0.5cm}
\caption{(Color online) Schematic diagram of the thermodynamic
processes of the QSZE with an isothermal expansion. For simplicity,
we only take two lowest energy levels with odd and even parities as
an example to show the rules of energy level redistributions due to
the insertion. (a) Initially a
single particle is in a thermal equilibrium with a heat bath at temperature $%
T$. (b) After adiabatically adding an infinite potential barrier
at the center of square well, the well is split into two identical
subspaces and before the measurement the particle stays in the
left or right subspaces with the same probability. (c) The
particle is found in the left subspace after the measurement. (d)
The system contacts with the heat bath of temperature $T$ and
reaches thermal equilibrium. (e) The system is performed an
isothermal expansion. (f) The barrier arrives at the right end of
the well and the system returns to its initial state. }
\end{figure}
\begin{figure}[tbp]
\centering \includegraphics[width=12cm]{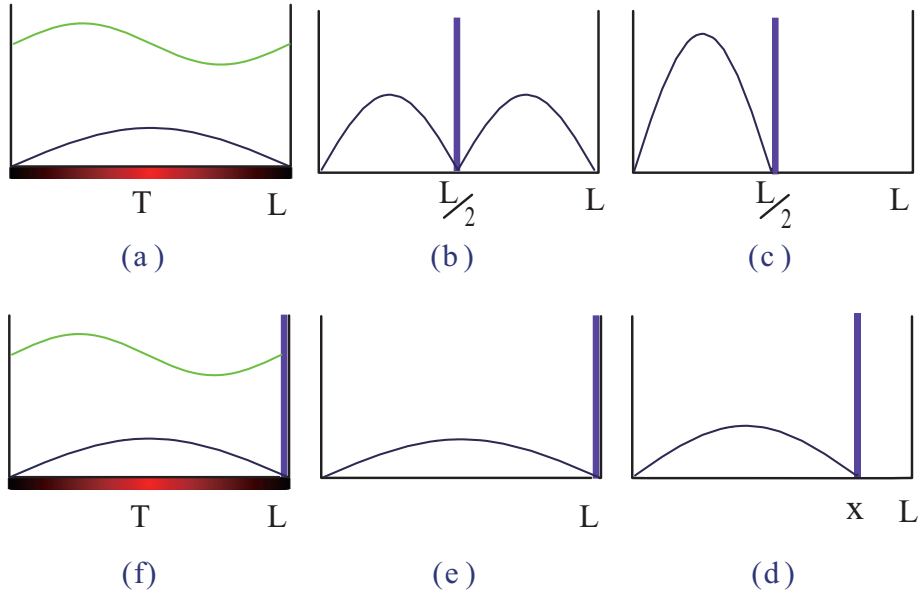} \hspace{0.5cm}
\caption{(Color online) Schematic diagram of the thermodynamic
processes of the QSZE with an adiabatic expansion. For the same
reason as in Fig. 2 we only take two lowest energy levels with odd
and even parities as an example. (a) Initially a single particle is
in thermal equilibrium with a heat bath at temperature $T$. (b)
After adiabatically adding an infinite potential barrier at the
center of square well, the well is split into two identical
subspaces and before the measurement the particle stays in the left
or right subspaces with the same probability. (c) The particle is
found in the left subspace after the measurement. (d) The system is
performed an adiabatic expansion. (e) The barrier arrives at the
right end of the well. (f) The system contacts with the heat bath
and reaches thermal equilibrium. }
\end{figure}


\section{The fully quantum analysis and discussions of the QSZE}\label{sec:quantum ana}

In this section we will present two different cyclic strategies,
one with isothermal expansion as illustrated in Fig. 2 and the
other with adiabatic expansion in Fig. 3. Each strategy consists
of four steps: adiabatic insertion, measurement, expansion and
removal. The first two steps, adiabatic insertion and measurement,
usually are performed simultaneously and can be considered as one.
We also assume that measurement is perfect, and the case of
imperfect measurement has been discussed in Ref.
\cite{[20]Sagawa}. In order to reveal the physics behind each
process, we calculate the internal energy, work, heat and the
entropy change in each step.


\subsection*{Step One: Adiabatic Insertion and Measurement}\label{subsec:one}

It is widely accepted that when a wall is inserted at any position
of the box, there is no heat and work accompanied by in the CSZE.
However, this is not the case in the QSZE. Analogous to the
classical insertion, the corresponding quantum process is
characterized by increasing the height of the potential barrier.
Adiabatic insertion $dQ=0$ implies that the system is isolated
from the heat bath and the potential barrier increases very slowly
at the center of potential well $x_{0}=L/2$. One can model the
potential
\begin{equation}\label{8}
V(x,t)  = \left
\{\begin{aligned} \infty,~~~~~~~~~~~~~~~~ &x<0,~ x>L \\
\lambda(t)\delta(x-\frac{L}{2}),~~~&0<x<L
 \end{aligned}
 \right.,
\end{equation}%
where $\lambda (t)$ varies from zero to infinity adiabatically. The
insertion of the impenetrable barrier is completed at $\lambda
(t)\rightarrow \infty .$


Appendix A shows the detailed calculations of the energy level
shifts during the adiabatic insertion (see also \cite{[Z]zhang}).
There are two
situations when the insertion takes place at $x_{0}=L/2$. The \emph{%
odd-parity} eigenfunctions $|E_{2k}(L)\rangle $ remain the same in
full space, because the insertion point is the same as nodes of
the odd-parity eigenfunctions such that the system cannot detect
and therefore does not resist the insertion. When the insertion is
completed, $|E_{2k}(L)\rangle $ become the eigenfunction
$|E_{k}(L/2)\rangle $ in the left and right subspaces with equal
probability. Since $E_{k}(L/2)=E_{2k}(L)$, the energy levels do
not shift in this situation.

The insertion changes the \emph{even-parity} eigenfunctions $%
|E_{2k-1}(L)\rangle $ and the eigenenergies. Interestingly, the
eigenenergies also vary with the insertion rate \cite{[14]Bender}.
When the
adiabatic insertion is completed, these eigenfunctions become $%
|E_{k}(L/2)\rangle $ in the left and right subspaces with equal
probability such that $E_{k}(L/2)=E_{2k}(L)$. \ This suggests that
$E_{2k-1}(L)$ will shift upward to the nearest level $E_{2k}(L)$.


Since the potential tends to infinity when the insertion is
completed, the box will be divided into two independent unrelated
subspaces, and the particle will be either in the left subspace or
the right subspace with the same probability,
$P^{(L)}=P^{(R)}=1/2$. So the cross terms of $\rho^{(L)}$ and
$\rho^{(R)}$, in the density matrix of the system $\rho_{1}$,
become zero after the insertion and $\rho_{1}$ reads as
\begin{equation}
\begin{array}{l}
\rho _{1}=\frac{1}{2}(\rho ^{(L)}+\rho
^{(R)}).%
\end{array}
\label{10}
\end{equation}%
The density matrices of subspaces $\rho ^{(L)}$ and $\rho ^{(R)}$
can be expressed as
\begin{equation}
\rho ^{(L)}=\sum_{k=1}^{\infty }P_{k}^{(L)}(\frac{L}{2})|E_{k}^{(L)}(\frac{L%
}{2})\rangle \langle E_{k}^{(L)}(\frac{L}{2})|,
\label{9}
\end{equation}%
\begin{equation}
\rho ^{(R)}=\sum_{k=1}^{\infty }P_{k}^{(R)}(\frac{L}{2})|E_{k}^{(R)}(\frac{L%
}{2})\rangle \langle E_{k}^{(R)}(\frac{L}{2})|,
\label{9'}
\end{equation}%
and
\begin{equation}
P_{k}^{(L)}(\frac{L}{2})=P_{k}^{(R)}(\frac{L}{2})=P_{2k}(L)+P_{2k-1}(L),
\label{a1}
\end{equation}%
\begin{equation}
E_{k}^{(L)}(\frac{L}{2})=E_{k}^{(R)}(\frac{L}{2})=E_{2k}(L),
\label{a2}
\end{equation}%
where $L$ and $R$ denote the left subspace and the right subspace.
The eigenstates $|E_{k}^{(L)}(L/2)\rangle $ and
$|E_{k}^{(R)}(L/2)\rangle $ correspond to the eigenvalues
$E_{k}^{(L)}(L/2)$ and $E_{k}^{(R)}(L/2)$ with the well width $L/2$.
$P_{k}^{(L)}(L/2)$ ( $P_{k}^{(R)}(L/2)$) is the
probability in the state $|E_{k}^{(L)}(L/2)\rangle $ ( $%
|E_{k}^{(R)}(L/2)\rangle $) immediately after the insertion is
completed. At
this moment the system is not yet in thermal equilibrium with the bath and $%
P_{k}^{(L)}(L/2)$ or $P_{k}^{(R)}(L/2)$ does not satisfy the
Boltzmann distribution. The entropy of the system is
\begin{equation}
S_{1}=-k_{B}\mathtt{Tr}(\rho _{1}\mathtt{ln}\rho _{1})=S_{c}+h(p),
\label{11}
\end{equation}%
where $S_{c}=k_{B}\mathtt{ln}2$, $h(p)=-k_{B}\mathtt{Tr}(\rho ^{(L)}\mathtt{%
ln}\rho ^{(L)})$ represent the classical information entropy and
quantum information entropy of the system respectively \ (we will
sometimes use information to refer to entropy in the following
context). The internal energy is
\begin{equation}
U_{1}=\frac{1}{2}\sum_{k=1}^{\infty }[P_{k}^{(L)}(\frac{L}{2})E_{k}^{(L)}(%
\frac{L}{2})+P_{k}^{(R)}(\frac{L}{2})E_{k}^{(R)}(\frac{L}{2})].%
\label{12}
\end{equation}%
Substitute Eqs. (\ref{a1}) and (\ref{a2}) into Eq. (\ref{12}) one
obtains
\begin{equation}
U_{1}=\sum_{k=1}^{\infty }[P_{2k}(L)+P_{2k-1}(L)]E_{2k}(L).
\label{13}
\end{equation}%
The internal energy change merely comes from the work done by the
outside agent because the insertion is implemented adiabatically,
i.e., $Q_{1}=0$. In general, the measurement
is performed without any energy cost \cite{[16]Kim,[14]Bender}, so the work $%
W_{1}$ done by the outside agent, in the insertion process, equals
the amount of the increased internal energy, $\Delta U_{10}$, that
is
\begin{equation}
W_{1}=\Delta U_{10}=U_{1}-U_{0}=\sum_{k=1}^{\infty
}P_{2k-1}(L)[E_{2k}(L)-E_{2k-1}(L)].  \label{14}
\end{equation}%
The total entropy change of the system is
\begin{equation}
\Delta S_{10}=S_{1}-S_{0}=S_{c}-[S_{0}-h(p)],  \label{15}
\end{equation}%
where $S_{0}-h(p)>0$ and can be easily verified through the inequality $%
P_{2k-1}\mathtt{ln}P_{2k-1}+P_{2k}\mathtt{ln}P_{2k}<(P_{2k-1}+P_{2k})\mathtt{%
ln}(P_{2k-1}+P_{2k})$. Interestingly, Eq. (\ref{15}) shows that
the information change in the insertion consists of two parts: the
increased
classical information $S_{c}$, and the decreased quantum information, $%
-[S_{0}-h(p)]$. This implies that in the insertion process the
classical information increases while the quantum information
decreases.
In addition, it is noted that though the heat exchange is zero in
the adiabatic insertion, $dQ=0$, the entropy change of the system,
$\Delta S_{10}$, is not equal to zero because this process is a
non-equilibrium process, and the relation $dQ=TdS$ doesn't hold
any more.


The classical information and the quantum information are acquired
from different origins. The former comes from the position
distribution of the particle, while the later comes from the
probability distribution of the energy levels in quantum system
which is obtained at the expense of the work done by the outside
agent. In the whole cycle, they also play different roles. The
classical information $S_{c}$ seems to behave like a feedback
controller determining the moving direction of the barrier and
does not contribute to the heat absorption, however, the quantum
information $S_{0}-h(p)$ determines the heat absorbed. We will
show the differences specifically in subsequent sections.

We make a measurement to localize the particle in one of two sides
of the well. After the measurement the classical information
becomes zero and the particle, we assume, is in the left side
(same discussions when it is in the right). Then the feedback is
finished. Based on the result of the feedback, the barrier will
eventually reach the right end of the well. The state $\rho _{1}$,
after measurement, collapses into $\rho _{2}=\rho ^{(L)}$ in the
left space. The internal energy of the system now becomes
\begin{equation}
U_{2}=\sum_{k=1}^{\infty }P_{k}^{(L)}(\frac{L}{2})E_{k}^{(L)}(\frac{L}{2})=U_{1}.
\label{16}
\end{equation}%
It is commonly accepted that there is no cost of energy in the
measurement process such that the heat absorbed and work done for
the system during measurement are zero
\begin{equation}
Q_{2}=W_{2}=0.  \label{17}
\end{equation}%
But the entropy of the system changes and becomes
\begin{equation}
S_{2}=-k_{B}\mathtt{Tr}(\rho _{2}\mathtt{ln}\rho _{2})=-k_{B}\mathtt{Tr}%
(\rho ^{(L)}\mathtt{ln}\rho ^{(L)})=h(p).  \label{18}
\end{equation}%
The entropy change due to the measurement is $\Delta
S_{21}=S_{2}-S_{1}=-S_{c}$. This indicates again that we now know
exactly the side where the particle is located and the classical
information disappears after the feedback.


\subsection*{Step Two: Expansion Process}

 We will discuss two cyclic strategies associated with different expansions. One is the isothermal expansion
described in Fig. 2. The other is that the system first undergoes
an adiabatic expansion, and then relaxes to thermal equilibrium by
contacting the heat bath as illustrated in Fig. 3.


\subsection*{Case (A) Isothermal Expansion}

The isothermal expansion consists of two procedures. We first
''hold'' the barrier, let the system contact the heat bath and
wait until they reach thermal equilibrium, as shown in Fig.
2(c)$\rightarrow $(d). There is no work done in this procedure.
Second we let the barrier move very slowly and eventually arrive
at the right end as shown in Fig. 2(d)$\rightarrow $(f). We
require that the second procedure be quasi-static such that the
system and the heat bath are always in thermal equilibrium. The
system state at position $x$, can be described by the density
matrix $\rho (x)$
\begin{equation}
\rho (x)=\sum_{k=1}^{\infty }P_{k}(x)|E_{k}(x)\rangle \langle E_{k}(x)|,%
\label{b1}
\end{equation}%
where $L/2\leq x\leq L$, $P_{k}(x)=\exp [-\beta E_{k}(x)]/Z(x)$
represents
the probability of the particle at $k$ energy level $E_{k}(x)=\frac{%
{k^{2}\hbar ^{2}\pi ^{2}}}{{2mx^{2}}}$, and $Z(x)=\sum_{k=1}^{\infty
}e^{-\beta E_{k}(x)}$ is the partition function.


The first procedure, as shown in Fig. 2(d), ends up with the density matrix $%
\rho _{3}=\rho (x=L/2),$
\begin{equation}
\rho _{3}=\sum_{k=1}^{\infty
}P_{k}(\frac{L}{2})|E_{k}(\frac{L}{2})\rangle \langle
E_{k}(\frac{L}{2})|.  \label{19}
\end{equation}%
The work $W_{3}$ and the internal energy $U_{3}$ are
\begin{equation}
W_{3}=0,  \label{20'}
\end{equation}%
\begin{equation}
U_{3}=\sum_{k=1}^{\infty }P_{k}(\frac{L}{2})E_{k}(\frac{L}{2}),
\label{20}
\end{equation}%
respectively. The heat absorbed, $Q_{3}$, equals the increase of the
internal energy
\begin{equation}
Q_{3}=U_{3}-U_{2}=\sum_{k=1}^{\infty }[P_{k}(\frac{L}{2})-P_{k}^{(L)}(\frac{L%
}{2})]E_{k}(\frac{L}{2}).  \label{21}
\end{equation}%
The entropy of the system is
\begin{equation}
S_{3}=-k_{B}\mathtt{Tr}(\rho _{3}\mathtt{ln}\rho
_{3})=-k_{B}\sum_{k=1}^{\infty }[P_{k}(\frac{L}{2})\mathtt{ln}P_{k}(\frac{L}{%
2})].  \label{22}
\end{equation}%
The entropy change $\Delta S_{32}$ in this procedure is
\begin{equation}
\Delta S_{32}=S_{3}-S_{2}=-k_{B}[\mathtt{Tr}(\rho _{3}\mathtt{ln}\rho _{3})-%
\mathtt{Tr}(\rho _{2}\mathtt{ln}\rho _{2})].  \label{22'}
\end{equation}%
It shows that the system state changes from $\rho _{2}$ to $\rho
_{3}$ by absorbing heat $Q_{3}$ to erase the quantum information
$\Delta S_{32}$.


In the second procedure, the system entropy increases gradually and
reaches its maximum $S_{4}$ at $x=L$ where the system returns to the
initial equilibrium state $\rho _{0}$ so that
\begin{equation}
\rho _{4}=\rho _{0},  \label{25}
\end{equation}%
\begin{equation}
U_{4}=U_{0}(L),  \label{26}
\end{equation}%
\begin{equation}
S_{4}=S_{0}.  \label{27}
\end{equation}%
The quantum entropy change is given by
\begin{equation}
\Delta S_{43}=S_{4}-S_{3}=-k_{B}[\mathtt{Tr}(\rho _{0}\mathtt{ln}\rho _{0})-%
\mathtt{Tr}(\rho _{3}\mathtt{ln}\rho _{3})].  \label{28}
\end{equation}%

In the spirit of the Landauer's erasure principle in the CSZE, the
classical information is erased gradually with the moving barrier
and vanishes when the barrier reaches the end of the well. By
contrast the quantum information in the QSZE with the isothermal
expansion is erased gradually with the barrier moving until the
system returns to the initial state. The absorbed heat $Q_{4}$ has
erased the quantum information $\Delta S_{43}$. In this way all
the quantum information obtained in the insertion and the
measurement has been erased completely,
\begin{equation}
S_{0}-h(p)-[\Delta S_{32}+\Delta S_{43}]=0.  \label{28'}
\end{equation}%
This indicates that the amount of quantum information obtained
equals to the total amount of quantum information erased by
absorbing heat in the cycle.


We now find out how much heat is absorbed and how much work is
done by the system in the second procedure, by using the first law
of thermodynamics $dQ=dU+dW$. Here $dQ$ and $dW$ are the heat
absorbed and the work done by the system, respectively
\cite{[11]Kieu,[12]Esposito},
\begin{equation}
dW=-\sum_{n}P_{n}dE_{n},  \label{29}
\end{equation}%
and
\begin{equation}
dQ=\sum_{n}E_{n}dP_{n}.  \label{30}
\end{equation}%
For slowly moving barrier, we can integrate Eqs. (\ref{29}) and
(\ref{30}) from $L/2$ to $L$ and obtain the work done by the
system and the heat absorbed in the isothermal expansion,
\begin{equation}
W_{4}=\sum_{k=1}^{\infty }\int_{L/2}^{L}P_{k}(x)dE_{k}(x)
=-\sum_{k=1}^{\infty }\int_{L/2}^{L}\frac{e^{-\beta E_{k}(x)}}{Z(x)}dE_{k}(x)%
=k_{B}T\mathtt{ln}\frac{Z(L)}{Z(L/2)},%
\label{36}
\end{equation}%

\begin{equation}
\begin{aligned}
Q_{4}=&\int_{L/2}^{L}d[U(x)+W(x)]
=U_{4}-U_{3}+k_{B}T\mathtt{ln}\frac{Z(L)}{Z(L/2)} \\
=&\sum_{k=1}^{\infty
}\{[P_{2k}(L)E_{2k}(L)+P_{2k-1}(L)E_{2k-1}(L)]-P_{k}(\frac{L}{2})E_{k}(\frac{L}{2})\}+k_{B}T\mathtt{ln}\frac{Z(L)}{
Z(L/2)}.
\end{aligned}
\label{35}
\end{equation}%
Here, the absorbed heat $Q_{4}$ is exploited to erase the amount
of the quantum information $\Delta S_{43}$ and brings the system
back to the initial thermal equilibrium state $\rho _{0}$ from
$\rho _{3}$.


The total work $W_{\mathtt{exp}}$ and the total heat
$Q_{\mathtt{exp}}$ in the two procedures are
\begin{equation}
W_{\mathtt{exp}}=W_{3}+W_{4}=k_{B}T\mathtt{ln}\frac{Z(L)}{Z({L/2})},
\label{36}
\end{equation}%
\begin{equation}
\begin{aligned}
Q_{\mathtt{exp}}=&Q_{3}+Q_{4}
=-\sum_{k=1}^{\infty }P_{2k-1}(L)[E_{2k}(L)-E_{2k-1}(L)]+k_{B}T%
\mathtt{ln}\frac{Z(L)}{Z({L/2})}
=-W_{1}+W_{\mathtt{exp}},%
\end{aligned}
\label{37}
\end{equation}%
respectively. Eq. (\ref{37}) can also be rewritten as $Q_{\mathtt{exp}%
}+W_{1}=W_{\mathtt{exp}}$. It implies that the total heat plus the
work done by the external agent in the insertion,
$Q_{\mathtt{exp}}+W_{1}$, equals to the work $W_{\mathtt{exp}}$ done
by the system in the expansion. Moreover, the quantum information
$S_{0}-h(p)$ is used to extract heat from the bath and determines
how much heat is absorbed or how much work is extracted.


\subsection*{Case (B) Adiabatic Expansion and Thermalization}

Similar to the strategy of the isothermal expansion, the process
also consists of two procedures. The system first undergoes an
adiabatic expansion to the right end of square well, and then
contacts the heat bath and relaxes to thermal equilibrium.


 The absorbed heat $Q_{3}^{\prime }=0$ in the adiabatic expansion, as
illustrated in Fig. 3(c)$\rightarrow $(e). The process is
non-equilibrium, and during the process the probability
distribution of the energy level is the same as that immediately
after the insertion and measurement. Let $P_{k}(x)$
be the probability at the energy level $k$ and at position $x$, then $%
P_{k}(x)=P_{k}^{(L)}(\frac{L}{2})$, where $\frac{L}{2}\leq x\leq L$ and $%
P_{k}^{(L)}(\frac{L}{2})$ is given in Eq. (\ref{a1}). When the
barrier reaches the end of well at $x=L$, the internal energy
$U_{3}^{\prime }(L)$ can be written as
\begin{equation}
U_{3}^{\prime }=\sum_{k=1}^{\infty
}P_{k}^{(L)}(\frac{L}{2})E_{k}(L), \label{38}
\end{equation}%
and the quantum entropy of the system does not change,
\begin{equation}
S_{3}^{\prime }=S_{2}=h(p).  \label{39}
\end{equation}%
Since no heat is absorbed in this process, the work $W_{3}^{\prime
}$ equals to the internal energy decrease,
\begin{equation}
\ W_{3}^{\prime }=U_{2}-U_{3}^{\prime }
=\sum_{k=1}^{\infty}P_{k}^{(L)}(\frac{L}{2})[E_{k}(\frac{L}{2})-E_{k}(L)]
=\sum_{k=1}^{\infty }3[P_{2k}(L)+P_{2k-1}(L)]E_{k}(L).%
\label{3a}
\end{equation}


The system afterwards contacts the heat bath and relaxes to thermal
equilibrium (i.e., thermalization), as described in Fig. 3(e)$\rightarrow $%
(f). There is no work done in this process, i.e., $W_{4}^{\prime
}=0$. The system only absorbs heat $Q_{4}^{\prime }$,
\begin{equation}
\begin{aligned}
Q_{4}^{\prime }=&U_{0}-U_{3}^{\prime } \\
=&\sum_{k=1}^{\infty }[P_{2k}(L)E_{2k}(L)+P_{2k-1}(L)E_{2k-1}(L)]
-\sum_{k=1}^{\infty }P_{k}^{(L)}(\frac{L}{2})E_{k}(L) \\
=&\sum_{k=1}^{\infty }3[P_{2k}(L)+P_{2k-1}(L)]E_{k}(L)-P_{2k-1}(L)[E_{2k}(L)-E_{2k-1}(L)] \\
=&W_{3}^{\prime }-W_{1},%
\end{aligned}
\label{3b}
\end{equation}%
and it is clear that $Q_{4}^{\prime }$ is also the total heat
absorbed in a cycle with the adiabatic expansion and is exploited to
erase the quantum information $\Delta S_{0}-h(p)$.


The above analysis shows that the total heat
$Q_{\mathtt{exp}}^{\prime }$ and the work $W_{\mathtt{exp}}^{\prime
}$ in this strategy are
\begin{equation}
Q_{\mathtt{exp}}^{\prime }=Q_{3}^{\prime }+Q_{4}^{\prime
}=Q_{4}^{\prime }, \label{4a}
\end{equation}%
\begin{equation}
W_{\mathtt{exp}}^{\prime }=W_{3}^{\prime }+W_{4}^{\prime
}=W_{3}^{\prime }, \label{4b}
\end{equation}%
respectively. It turns out that the heat $Q_{\mathtt{exp}}^{\prime
}$ is the same as the heat $Q_{4}^{\prime }$. Eqs. (\ref{3b}) -
(\ref{4b}) demonstrate that the absorbed heat and the work done for
the system in this strategy
satisfy the relation $Q_{\mathtt{exp}}^{\prime }+W_{1}=W_{\mathtt{exp}%
}^{\prime }$. This is the same as that of the isothermal expansion
discussed in case (A) (Eq. (\ref{37})).

However, it is interesting to note that although the quantum
information is the same in both strategies, the results of the
work and the heat are different. The heat absorbed and the work
done in the QSZE depend on the cyclic strategies, that are
deviated from $k_{B}T\mathtt{ln}2$ in the CSZE.

\subsection*{Step Three: Removal Process}
 As mentioned above, the barrier will
always end up at the edge of the well when the expansion is
completed. Since the system will not be disturbed by
removing the barrier, the work and heat are zero, $W_{\mathtt{rem}}=0$ and $%
Q_{\mathtt{rem}}=0$. After the removal the system returns to its
initial state. We have analyzed each step of the whole cycle, and
now let's check out the relations between the total heat absorbed
and the total net work done by the system for each strategy.

For the isothermal case, the total amount of heat $Q_{\mathtt{tot}}$
absorbed from the heat bath is
\begin{equation}
Q_{\mathtt{tot}}= Q_{1}+Q_{2}+Q_{3}+Q_{4}= Q_{\mathtt{exp}}
=-\sum_{k=1}^{\infty
}P_{2k-1}(L)[E_{2k}(L)-E_{2k-1}(L)]+k_{B}T\mathtt{ln}\frac{Z(L)}{Z(\frac{L}{2})}.
\label{41}
\end{equation}%
It shows that $Q_{\mathtt{tot}}$ helps to recover the quantum
entropy to its maximum value $S_{0}$ from $h({p})$. In another
word, the quantum information $S_{0}-h({p})$ can help the system
absorb heat, $Q_{\mathtt{tot}}$, and eventually the system returns
to the initial state. The insertion work $W_{\mathtt{ins}}$ equals
to the minus $W_{1}$, $W_{\mathtt{ins}}=-W_{1}$, such that the
total work $W_{\mathtt{tot}}$ and total heat $Q_{\mathtt{tot}}$
can be expressed as
\begin{equation}
W_{\mathtt{tot}}=W_{\mathtt{ins}}+W_{\mathtt{exp}}+W_{\mathtt{rem}}=Q_{%
\mathtt{tot}}=-\sum_{k=1}^{\infty
}P_{2k-1}(L)[E_{2k}(L)-E_{2k-1}(L)]+k_{B}T\mathtt{ln}\frac{Z(L)}{Z(\frac{L}{2})}.
\label{42}
\end{equation}%
It suggests that the total absorbed heat is fully transformed into
the effective work of the system, which brings the system into the
initial state.


The adiabatic expansion is similar to the isothermal. The total heat $Q_{%
\mathtt{tot}}^{\prime }$ and total work $W_{\mathtt{tot}}^{\prime }$
satisfy
\begin{equation}
Q_{\mathtt{tot}}^{\prime }=W_{\mathtt{\mathtt{tot}}}^{\prime }
=\sum_{k=1}^{\infty }3[P_{2k}(L)+P_{2k-1}(L)]E_{k}(L)-P_{2k-1}(L)[E_{2k}(L)-E_{2k-1}(L)].%
\label{44}
\end{equation}%
Eqs. (\ref{42}) and (\ref{44}) show that the total amount of heat
absorbed equals to the total work (net work) for both of cyclic
strategies.


We now summarize the above discussions as follows:
\\(a) \begin{minipage}[t]{452pt}
We consider the four physical quantities, internal energy, work,
heat and entropy (information) in both CSZE and QSZE, and compare
their physical properties. In the CSZE the internal energy is
conserved during the whole cycle, while in the QSZE the internal
energy is changed in the insertion process and in both the
isothermal and adiabatic expansion processes. The CSZE only does
work in the expansion process during the entire cycle, while in
the QSZE work is done in both insertion and expansion processes.
In the whole cycle of the CSZE the system only absorbs heat in the
expansion process, while in the QSZE it is not that case. In the
strategy with the isothermal expansion the heat exchange between
the system and the bath occurs twice: After the insertion
contacting with the bath and the expansion process, and in the
strategy with the adiabatic expansion it occurs only after the
expansion contacting the bath. The total absorbed heat is fully
exploited to erase the quantum information, while in the CSZE the
absorbed heat is only used to erase the classical information.
This paper classifies the entropy into the classical entropy and
the quantum information entropy. They have different origins. The
former reflects the distribution of the particle's position and is
the same as that in the classical system. The later is determined
by the probability distribution of the energy levels. In the QSZE,
they play entirely different roles. The classical information
seems to behave like a feedback controller and has no contribution
to extracting heat from the bath, while the quantum information
acquired during the insertion determines the amount of heat
absorbed and work done for the system in a cycle.\end{minipage}
\\(b) \begin{minipage}[t]{452pt}The information itself cannot be
converted into energy but it could be exploited to extract work or
heat \cite{[22]Lev B} and, in the QSZE, the quantum information
determines the amount of heat absorbed and work done by the
system.
\end{minipage}
\\(c) \begin{minipage}[t]{452pt}
Since the insertion and measurement lead to the quantum entropy
decrease, the heat must be required, in subsequent processes, to
compensate the quantum entropy change and brings the system to the
initial state. This is consistent with the spirit of Landauer's
information erasing principle \cite{[4]Landauer}. In the original
ideal CSEZ the insertion does not need work such that the whole
process results in extracting the energy $k_{B}T\mathtt{ln}2$ from
the bath or the entropy decrease $k_{B}\mathtt{ln}2$. On the
contrary, the net effect in the QSEZ is that the system absorbs
heat from the bath, obtains work in the insertion process, and
does work to the outside during the expanding process. It is noted
that the work obtained in the insertion process is not the same as
that lost in the expansion process.  Therefore, although the
system returns to the initial state after a cycle, the outside
world will not return to its initial state. It turns out that the
second law of thermodynamics is not violated in the
QSZE.\end{minipage}

\section{Discussions of two limits}
Now let's focus on the two limits $L\rightarrow \infty $ and
$T\rightarrow \infty $.


\emph{Case (A)} $L\rightarrow \infty $:

The partition function $Z(L)$ in the limit $L\rightarrow \infty $
is
\begin{equation}
Z(L)=\sum_{k=1}^{\infty }\mathtt{exp}(-\xi k^{2}),  \label{45}
\end{equation}%
where $\xi =\frac{1}{k_{B}T}\frac{\pi ^{2}\hbar ^{2}}{2mL^{2}}>0$. When $%
L\rightarrow \infty $, the parameter $\xi $ goes to zero and the sum
in the above expression can be replaced by an integral
\begin{equation}
Z(L)=\int_{1}^{\infty }\mathtt{\ exp}(-\xi x^{2})dx\approx \frac{\sqrt{\pi }%
}{2}\xi ^{-\frac{1}{2}}=\frac{L}{\sqrt{h^{2}/{2\pi mk_{B}T}}},
\label{46}
\end{equation}%
such that
\begin{equation}
\lim_{L\rightarrow \infty }W_{\mathtt{exp}}=\lim_{L\rightarrow \infty }k_{B}T%
\mathtt{ln}\frac{Z(L)}{Z(\frac{L}{2})}=k_{B}T\mathtt{ln}2.
\label{46'}
\end{equation}%
Based on Eqs. (\ref{B5}) and (\ref{B11}) in Appendix B, we obtain
\begin{equation}
\lim_{L\rightarrow \infty }W_{1}=0,  \label{47}
\end{equation}%
\begin{equation}
\lim_{L\rightarrow \infty }\frac{W_{1}}{U_{0}}=0.  \label{48}
\end{equation}%
The two expressions suggest that the work in the insertion goes to
zero when $L\rightarrow \infty $, and it hardly has influence on the
system's internal
energy. According to Eq. (\ref{B13}), we have $\lim_{L\rightarrow \infty }%
\frac{W_{1}}{W_{\mathtt{exp}}}=0$, meaning that $W_{1}$ is much
smaller than $W_{\mathtt{exp}}$ and can be ignored in the total
work. We therefore obtain
\begin{equation}
W_{\mathtt{tot}}=W_{\mathtt{ins}}+W_{\mathtt{exp}}+W_{\mathtt{rem}}=k_{B}T%
\mathtt{ln}\frac{Z(L)}{Z(\frac{L}{2})}\approx k_{B}T\mathtt{ln}2.
\label{49}
\end{equation}%
Eq. (\ref{C8}) in Appendix C shows that when $L\rightarrow \infty
$, the quantum information $S_{0}-h(p)$ acquired in the insertion
will reduce to the classical information $k_{B}\mathtt{ln}2$. From
Eq. (\ref{15}) the total information change in the insertion will
become zero. From above discussions we now arrived at the
conclusion: When $L\rightarrow \infty $, none of interested four
physical quantities changes in the insertion process. It turns out
that the QSZE reduces to the CSZE completely.


\emph{Case (B)} $T\rightarrow \infty $:

Based on the definition $\xi =\frac{1%
}{k_{B}T}\frac{\pi ^{2}\hbar ^{2}}{2mL^{2}}$, we can show that $%
\lim_{T\rightarrow \infty }\xi =\lim_{L\rightarrow \infty }\xi
=0$. Then in high-temperature limit, the partition function $Z(L)$
is the same as
Eq. (\ref{46}). We also have $\lim_{T\rightarrow \infty }W_{\mathtt{exp}%
}=k_{B}T\mathtt{ln}2$ which is the same as that in the
$L\rightarrow \infty $ case. Although the inserting work $W_{1}$
is divergent \cite{[22]Dong}, Eqs. (\ref{B14}, \ref{B15}) in
Appendix B tell us that when $T\rightarrow \infty $, the internal
energy is hardly changed in the insertion because the work by
the external agent is much less than the internal energy, such that%
\begin{equation}
W_{\mathtt{tot}}=W_{\mathtt{ins}}+W_{\mathtt{exp}}+W_{\mathtt{rem}}\approx
k_{B}T\mathtt{ln}\frac{Z(L)}{Z(\frac{L}{2})}=k_{B}T\mathtt{ln}2.
\label{53}
\end{equation}%
Eq. (\ref{C8}) in Appendix C shows that the quantum information
$S_{0}-h(p)$ acquired in the insertion will also reduce to the classical information $%
k_{B}\mathtt{ln}2$ and the total information change in the insertion
becomes zero.


From the above discussions it is clear that the QSZE will reduce
to the CSZE in $L\rightarrow \infty $ and $T\rightarrow \infty $
limits, that is just the result what we expected.


\section{Conclusions}

To summarize, we gave the detailed analysis and discussions on the
QSZE of a single particle confined to a one-dimensional infinite
square well with fully quantum consideration. We for the first
time considered the energy level shifts in the insertion, and
investigated its effect on physical quantities, such as heat,
work, internal energy, and entropy. We found that only the quantum
information contributes to the absorbed heat, while the classical
information acts like a feedback controller and has no direct
effect on the heat absorbed from bath. We also demonstrated that
the
work done by the system is different from $W_{\mathtt{tot}}=k_{B}T\mathtt{ln%
}2$. It is noted that unlike in the CSZE the external agent in the
QSZE has to do some work in the insertion process, and the one
does the work in the insertion process is not the same one to
which the system does the work in the expansion process. Although
the system returns to the initial states after one cycle, the
outside world will not return to its initial state. The second law
of thermodynamics therefore is not violated in the QSZE. In the
limits of $L\rightarrow \infty $ or $T\rightarrow \infty $, the
quantum Szilard engine (QSZE) reduces to the classical Szilard
engine (CSZE), and the relation
$W_{\mathtt{tot}}=k_{B}T\mathtt{ln}2$ holds again. Significantly,
it is the first demonstration of the different effects between
quantum information and classical information for extracting heat
from the bath in the QSZE, which provides further understanding of
the relationships among heat, information and work from
quantum-mechanical perspective.

\bigskip{}
\section*{Acknowledgements}
This work is financially supported by National Science Foundation
of China (Grants No. 10974016, 11075013, and 11005008), the
Natural Science Foundation of Shandong Province, China (Grant No.
ZR2011FL009) and the Science and Technology Project of University
in Shandong Province, China (Grant No. J12LJ01). L. -A. Wu has
been supported by the Ikerbasque Foundation Start-up, the Basque
Government (grant IT472-10) and the Spanish MEC (Project
No.FIS2009-12773-C02-02).


\appendix

\label{appendix1}

\section{Derivation of the rule of energy level redistribution in the limit of the
height of the barrier tending to infinity}


We use $\psi (x)$ instead of $|E(x)\rangle $ to denote the wave
function of the system for simplicity. For a single particle of mass
$m$ in a potential field described by Eq. (\ref{8}), $\psi (x)=0$,
in the $x<0$ and $L<x$ regions. When $0<x<L$, $\psi (x)$ satisfies
Schr\"{o}dinger equation

\begin{equation}\label{f1}
-\frac{\hbar ^{2}}{2m}\frac{d^{2}\psi (x)}{dx^{2}}+\lambda \delta (x-\frac{L%
}{2})\psi (x)=E\psi (x),
\end{equation}
and the boundary conditions are
\begin{equation}\label{f2}
 \psi(0)=0,~~~   \psi(L)=0 ,~~~ \psi(\frac{L^+}{2})=\psi(\frac{L^-}{2}) ,
\end{equation}
\begin{equation}\label{f3}
\psi'(\frac{L^+}{2})-\psi'(\frac{L^-}{2})=\frac{2m\lambda}{\hbar^2}\psi(\frac{L}{2}).
\end{equation}
The general solution of Eq. (\ref{f1}) is
\begin{equation} \label{eq:1}
\left\{ \begin{aligned}
         \psi_1(x)=A\sin (kx+\varphi_1),~~~~ 0<x<\frac{L}{2} \\
                  \psi_2(x)=B\sin (kx+\varphi_2) ,~~~\frac{L}{2}<x<L
                          \end{aligned} \right.,
                          \end{equation}
where $k=\sqrt{\frac{2mE}{\hbar ^{2}}}$. Eq. (\ref{eq:1}) represents
the eigenfunctions of the system. We will check that, for the
stationary wave
functions $\psi _{n}(x)=\sqrt{\frac{2}{L}}\sin \frac{n\pi x}{L}%
,~~~n=1,2,3\cdots $, when the integer n is even $n=2,4,6,\cdots $,
the wave function $\psi _{n}(\frac{L}{2})=0$ holds, and all the
boundary conditions are satisfied. The solutions of Eq. (\ref{f1})
with even $n$ are
\begin{equation}
\psi _{n}(x)=\sqrt{\frac{2}{L}}\sin \frac{n\pi
x}{L},~~~n=2,4,6\cdots . \label{f4}
\end{equation}%
For the other solutions, we substitute Eq. (\ref{eq:1}) into Eq.
(\ref{f3}) and have $\varphi _{1}=0$, $\varphi _{2}=-kL$, and
$B=-A$.  Eq. (\ref{eq:1}) can be expressed as
\begin{equation}
\left\{ \begin{aligned} \psi_1(x)=A \sin kx,~~~~~~~~~~~ 0<x<\frac{L}{2} \\
\psi_2(x)=A\sin k(L-x) ,~~~\frac{L}{2}<x<L \end{aligned}\right..
\label{eq:2}
\end{equation}%
Substitute Eq. (\ref{eq:2}) into Eq. (\ref{f3}), one obtains
\begin{equation}
-\xi \cot \xi =\frac{mL}{2\hbar ^{2}}\lambda ,  \label{f5}
\end{equation}%
where $\xi =\frac{kL}{2}$. Eq. (\ref{f5}) indicates that the function $%
y=-\xi \cot \xi \geq 0$ is a periodic and monotone increasing
function in a period of $\frac{\pi }{2}$. For the arbitrary $i_{th}$
period, the function
becomes $y=-\xi _{i}\cot \xi _{i}=\frac{mL}{2\hbar ^{2}}\lambda $, where $%
\xi _{i}=\frac{k_{i}L}{2}=\frac{L}{2}\sqrt{\frac{2mE_{i}^{\prime
}}{\hbar
^{2}}}$, the variable $\xi _{i}$ satisfies the relation ($i-\frac{1}{2}%
)\pi <\xi _{i}<i\pi $, and the corresponding value of the function
$y$ varies from zero to infinity, namely, the parameter $\lambda $
changes from zero to infinity continuously. We therefore obtain
\begin{equation}
\frac{{(2i-1)}^{2}{\pi }^{2}\hbar ^{2}}{2mL^{2}}<E_{i}^{\prime }<\frac{{(2i)}%
^{2}{\pi }^{2}\hbar ^{2}}{2mL^{2}}.  \label{f7}
\end{equation}%
Eq. (\ref{f7}) can be also expressed as $E_{2i-1}<E_{i}^{\prime
}<E_{2i}$ which is equivalent to the inequality ($i-\frac{1}{2})\pi
<\xi _{i}<i\pi $.
The two sides of the inequality correspond to $\lambda \rightarrow 0$ and $%
\lambda \rightarrow \infty $ respectively. From Eq. (\ref{f4}) and Eq. (\ref%
{f7}), we conclude that the even levels $E_{2i}$ don't shift and the
odd levels $E_{2i-1}$ shift upwards to $E_{2i}$ when $\lambda $ goes
to infinity.


\label{appendix2}

\section{The work done by the external agent $W_{\mathtt{1}}$ in two limit cases}
~\\\emph{Case (A): $L \rightarrow \infty$}.\newline

According to Eq. (\ref{14}), $W_{1}$ can be written as
\begin{equation}
W_{1}=\frac{\sum_{k=1}^{\infty }P_{2k-1}(L)[E_{2k}(L)-E_{2k-1}(L)]}{%
\sum_{k=1}^{\infty }P_{k}}
=\frac{\sum_{k=1}^{\infty }\mathtt{exp}[-\xi (2k-1)^{2}](4k-1)\xi }{%
\beta \sum_{k=1}^{\infty }\mathtt{exp}(-\xi k^{2})},%
\label{B1}
\end{equation}%
where $\sum_{n=1}^{\infty }P_{n}(L)=1$, $P_{n}(L)=\frac{{e^{-\beta
E_{n}}}}{{Z(L)}}$, $Z(L)=\sum_{n=1}^{\infty }e^{-\beta E_{n}}$ is
the partition function and $\xi =\frac{1}{k_{B}T}\frac{\pi ^{2}\hbar
^{2}}{2mL^{2}}>0$. In view of the following relations:
\begin{equation}
\int_{1}^{\infty }e^{-\xi x^{2}}xdx<\sum_{k=1}^{\infty }e^{-\xi
k^{2}}k<\int_{0}^{\infty }e^{-\xi x^{2}}xdx,  \label{B2}
\end{equation}%
and
\begin{equation}
0<\int_{0}^{1}e^{-\xi x^{2}}dx<\int_{0}^{1}dx=1,  \label{B3}
\end{equation}%
we have
\begin{equation}
\begin{aligned}
W_{1}<&\frac{1}{\beta }\frac{4\int_{0}^{\infty }\xi e^{-\xi x^{2}}xdx}{%
\int_{1}^{\infty }e^{-\xi x^{2}}dx}\\
=&\xi \frac{1}{\beta }\frac{4\int_{0}^{\infty }e^{-\xi x^{2}}xdx}{%
\int_{0}^{\infty }e^{-\xi x^{2}}dx-\int_{0}^{1}e^{-\xi x^{2}}dx}\\
<&\xi \frac{1}{\beta }\frac{4\int_{0}^{\infty }e^{-\xi x^{2}}xdx}{%
\int_{0}^{\infty }e^{-\xi x^{2}}dx-\int_{0}^{1}dx}\\
=&\frac{1}{\beta }\frac{2}{\frac{\sqrt{\pi }}{2}\xi ^{\frac{-1}{2}}-1},%
\end{aligned}
\label{B4}
\end{equation}%
where $\beta =\frac{1}{k_{B}T}$ is a constant.
\begin{equation}
\lim_{L\rightarrow \infty }W_{1}=\lim_{\xi \rightarrow
0}W_{1}=\lim_{\xi
\rightarrow 0}\frac{1}{\beta }\frac{2}{\frac{\sqrt{\pi }}{2}\xi ^{\frac{-1}{2%
}}-1}=0.  \label{B5}
\end{equation}%
In the derivation of Eq. (\ref{B4}) we have used the following formula \cite%
{[p]Ralph}
\begin{equation}
I(n,a)=\int_{0}^{\infty }e^{-ax^{2}}x^{n}dx,  \label{B6}
\end{equation}%
where $a>0$, $n$ is an integer and $I(0,a)=\frac{\sqrt{\pi
}}{2}a^{-\frac{1}{2}} $, $I(1,a)=\frac{1}{2}a^{-1}$. By using the
same formula as Eq. (\ref{B6}) and the following relations:
\begin{equation}
\int_{1}^{\infty }e^{-\xi x^{2}}x^{2}dx<\sum_{k=1}^{\infty }e^{-\xi
k^{2}}k^{2}<\int_{0}^{\infty }e^{-\xi x^{2}}x^{2}dx,  \label{B8}
\end{equation}%
and
\begin{equation}
0<\int_{0}^{1}e^{-\xi x^{2}}x^{2}dx<\int_{0}^{1}x^{2}dx=\frac{1}{3},
\label{B9}
\end{equation}%
we have
\begin{equation}
\begin{aligned}
\frac{W_{1}}{%
U_{0}}=&\frac{\sum_{k=1}^{\infty }P_{2k-1}(L)[E_{2k}(L)-E_{2k-1}(L)]}{%
\sum_{k=1}^{\infty }P_{k}E_{k}(L)}\\
=&\frac{\sum_{k=1}^{\infty }\mathtt{exp}[-\xi (2k-1)^{2}](4k-1)}{%
\sum_{k=1}^{\infty }\mathtt{exp}(-\xi k^{2})k^{2}}\\
<&\frac{4\int_{0}^{\infty }e^{-\xi x^{2}}xdx}{\int_{1}^{\infty
}e^{-\xi x^{2}}x^{2}dx}\\
=&\frac{4\int_{0}^{\infty }e^{-\xi x^{2}}xdx}{\int_{0}^{\infty
}e^{-\xi x^{2}}x^{2}dx-\int_{0}^{1}e^{-\xi x^{2}}x^{2}dx} \\
=&\frac{2}{\frac{\sqrt{\pi }}{4}\xi ^{-\frac{1}{2}}-\frac{1}{3}\xi }.%
\end{aligned}
\label{B10}
\end{equation}%
Thus the limit is
\begin{equation}
\lim_{L\rightarrow \infty }\frac{W_{1}}{U_{0}}=\lim_{\xi \rightarrow
0}\frac{W_{1}}{U_{0}}=0.  \label{B11}
\end{equation}%

 From Eqs. (\ref{B5}) and (\ref{46'}) we have
\begin{equation}
\lim_{L\rightarrow \infty }\frac{W_{1}}{W_{\mathtt{exp}}}=0.
\label{B13}
\end{equation}
~\\\emph{Case (B): $T \rightarrow \infty$}.\newline

 Since the limit
\begin{equation}  \label{B14'}
\lim_{T\rightarrow \infty}\xi=\lim_{L\rightarrow \infty}\xi=0
\end{equation}
holds, for $T \rightarrow \infty$, similarly one can obtain the same
results as those in the \emph{Case (A)}, i.e.,
\begin{equation}  \label{B14}
\lim_{T\rightarrow \infty}\frac{W_{1}}{U_0}=\lim_{\xi\rightarrow
0}\frac{W_{1}}{U_0}=0,
\end{equation}
and
\begin{equation}  \label{B15}
\lim_{T\rightarrow \infty} \frac{W_{1}}{W_{\mathtt{exp}}}=\lim_{\xi%
\rightarrow 0} \frac{W_{1}}{W_{\mathtt{exp}}}=0.
\end{equation}


\label{appendix3}

\section{The entropy change $S_{0}-h(p)$ in two limit cases}

Here we consider $S_{0}-h(p)$ in the two limits ${T\rightarrow
\infty }$ and ${L\rightarrow \infty }$. From the Eqs.
(\ref{7},\ref{a1},\ref{18}), $\Delta \equiv\frac{S_{0}-h(p)}{k_{B}}$
can be expressed as
\begin{equation}
\begin{aligned}
\Delta =&-\mathtt{Tr}(\rho _{0}\mathtt{ln}\rho _{0})+\mathtt{Tr}(\rho ^{(L)}%
\mathtt{ln}\rho ^{(L)}) \\
=&-\sum_{k=1}^{\infty }P_{k}\mathtt{ln}P_{k}+\sum_{k=1}^{\infty
}(P_{2k-1}+P_{2k})\mathtt{ln}(P_{2k-1}+P_{2k})]\\
=&-\sum_{k=1}^{\infty}(P_{2k}\mathtt{ln}P_{2k}+P_{2k-1}\mathtt{ln}P_{2k-1})+\sum_{k=1}^{\infty }(P_{2k-1}+P_{2k})\mathtt{ln}(P_{2k-1}+P_{2k}) \\
=&\sum_{k=1}^{\infty }P_{2k}\mathtt{ln}(1+\frac{P_{2k-1}}{P_{2k}}
)+\sum_{k=1}^{\infty }P_{2k-1}\mathtt{ln}(1+\frac{P_{2k}}{P_{2k-1}}) \\
=&\frac{\sum_{k=1}^{\infty }e^{-\xi (2k)^{2}}\mathtt{ln}[1+e^{\xi
(4k-1)}]+\sum_{k=1}^{\infty }e^{-\xi(2k-1)^{2}}\mathtt{ln}[1+e^{-\xi
(4k-1)}]}{\sum_{k=1}^{\infty }e^{-\xi k^{2}}},%
\end{aligned}
\label{C1}
\end{equation}%
where $\xi =\frac{1}{k_{B}T}\frac{\pi ^{2}\hbar ^{2}}{2mL^{2}}>0$,
and the above expression satisfies
\begin{equation}
\begin{aligned}
\frac{\sum_{k=1}^{\infty }e^{-\xi k^{2}}\mathtt{ln}[1+e^{-\xi (2k+1)}]}{%
\sum_{k=1}^{\infty }e^{-\xi k^{2}}}<\Delta <\frac{\sum_{k=1}^{\infty
}e^{-\xi k^{2}}\mathtt{ln}[1+e^{(-1)^{k}\xi
(2k-1)}]}{\sum_{k=1}^{\infty
}e^{-\xi k^{2}}}.%
\end{aligned}
\label{C2}
\end{equation}%
For simplicity, let $y_{L}$ and $y_{R}$ represent the left side
and right side of above inequality. The left side is
\\
\\
\begin{equation}
\begin{aligned}
y_{L}=&\frac{\sum_{k=1}^{\infty }e^{-\xi k^{2}}\mathtt{ln}[1+e^{-\xi (2k+1)}]%
}{\sum_{k=1}^{\infty }e^{-\xi k^{2}}} \\
>&\frac{\sum_{k=1}^{\infty }e^{-\xi k^{2}}\mathtt{ln}[2e^{-\xi (2k+1)}]}{%
\sum_{k=1}^{\infty }e^{-\xi k^{2}}} \\
=&\frac{\sum_{k=1}^{\infty }e^{-\xi
k^{2}}\mathtt{ln}2-\sum_{k=1}^{\infty
}e^{-\xi k^{2}}\xi (2k+1)}{\sum_{k=1}^{\infty }e^{-\xi k^{2}}} \\
=&\mathtt{ln}2-\frac{2\sum_{k=1}^{\infty }e^{-\xi k^{2}}\xi
k+\sum_{k=1}^{\infty }e^{-\xi k^{2}}\xi }{\sum_{k=1}^{\infty
}e^{-\xi k^{2}}}\\
>&\mathtt{ln}2-\frac{2\int_{0}^{\infty }e^{-\xi x^{2}}\xi
xdx+\int_{0}^{\infty }e^{-\xi x^{2}}\xi dx}{\int_{0}^{\infty
}e^{-\xi x^{2}}dx-\int_{0}^{1}dx} \\
=&\mathtt{ln}2-\frac{\frac{\sqrt{\pi }}{2}\xi ^{\frac{1}{2}}+1}{\frac{%
\sqrt{\pi }}{2}\xi ^{\frac{-1}{2}}-1},%
\end{aligned}
\label{C3}
\end{equation}%
and the limit satisfies
\begin{equation}
\lim_{\xi \rightarrow 0}y_{L}>\lim_{\xi \rightarrow 0}[\mathtt{ln}2-\frac{%
\frac{\sqrt{\pi }}{2}\xi ^{\frac{1}{2}}+1}{\frac{\sqrt{\pi }}{2}\xi ^{\frac{%
-1}{2}}-1}]=\mathtt{ln}2.  \label{C4}
\end{equation}%
For the right side in Eq. (\ref{C2}), we have
\begin{equation}
\begin{aligned}
y_{R}=&\frac{\sum_{k=1}^{\infty }e^{-\xi
k^{2}}\mathtt{ln}[1+e^{{(-1)}^{k}\xi
(2k-1)}]}{\sum_{k=1}^{\infty }e^{-\xi k^{2}}} \\
<&\frac{\sum_{k=1}^{\infty }e^{-\xi k^{2}}\mathtt{ln}(2e^{\xi (2k-1)})}{%
\sum_{k=1}^{\infty }e^{-\xi k^{2}}} \\
<&\frac{\sum_{k=1}^{\infty }e^{-\xi k^{2}}\mathtt{ln}2+2\sum_{k=1}^{%
\infty }e^{-\xi k^{2}}\xi k}{\sum_{k=1}^{\infty }e^{-\xi k^{2}}} \\
=&\mathtt{ln}2+\frac{2\sum_{k=1}^{\infty }e^{-\xi k^{2}}\xi k}{%
\sum_{k=1}^{\infty }e^{-\xi k^{2}}} \\
<&\mathtt{ln}2+\frac{2\int_{0}^{\infty }e^{-\xi x^{2}}\xi xdx}{%
\int_{0}^{\infty }e^{-\xi x^{2}}dx-\int_{0}^{1}dx} \\
=&\mathtt{ln}2+\frac{1}{\frac{\sqrt{\pi }}{2}\xi ^{\frac{-1}{2}}-1}.%
\end{aligned}
\label{C5}
\end{equation}%
Take the limit $\xi \rightarrow 0$ in Eq. (\ref{C5}), we have
\begin{equation}
\lim_{\xi \rightarrow 0}y_{R}<\lim_{\xi \rightarrow 0}[\mathtt{ln}2+\frac{1}{%
\frac{\sqrt{\pi }}{2}\xi ^{\frac{-1}{2}}-1}]=\mathtt{ln}2.
\label{C6}
\end{equation}%
From Eqs. ({C4}, {C6}), one obtains
\begin{equation}
\lim_{\xi \rightarrow 0}\Delta =\mathtt{ln}2.  \label{C7}
\end{equation}%
According to $\lim_{T\rightarrow \infty }\xi =\lim_{L\rightarrow
\infty }\xi =0$ and $\Delta =\frac{S_{0}-h(p)}{k_{B}}$, we obtain
\begin{equation}
\lim_{L \rightarrow \infty }[{S_{0}-h(p)}]=\lim_{T\rightarrow \infty }[{%
S_{0}-h(p)}]=\lim_{\xi \rightarrow 0}k_{B}\Delta =k_{B}\mathtt{ln}2.
\label{C8}
\end{equation}


\begin{thebibliography}{99}


\bibitem{[1]Leff} H. S. Leff and A. F. Rex,
\emph{Maxwell's Demon 2 : Entropy, Classical and Quantum
Information, Computing}, Institute of Physics, Bristol, 2003.

\bibitem{[2]Maruyama} K. Maruyama, F. Nori, and V. Vedral, Rev. Mod. Phys.
81 (2009) 1.

\bibitem{[3]Szilard} L. Szilard, Z. Phys. 53 (1929) 840.

\bibitem{[4]Landauer} R. Landauer, IBM J. Res. Dev. 5 (1961) 183.

\bibitem{[5]Bennett} C. H. Bennett, Int. J. Theor. Phys. 21 (1982) 905.


\bibitem{[p]L. B.1} L. B. Levitin, In: D. Cabile, D. G.
Kuper and I. Riess (eds.) Proc. 13th IUPAP Conf. Stat. Phys.
Hilger, Bristol (1978).

\bibitem{[p]L. B.2} L. B. Levitin, In: S. Diner, and G. Lochak, (eds.) \emph{Information, Complexity, and Control in Quantum
Physics}, pp. $15-47$. Springer, Berlin (1987).

\bibitem{[p]L. B.3} L. B. Levitin, In: \emph{Proc. Worksh. on Physics and Computation (PhysComp'92)}, pp. $223-226$.
IEEE Comput. Soc., Los Alamitos (1993).

\bibitem{[p]Scully} M. O. Scully, Phys. Rev. Lett. 87 (2001) 220601.

\bibitem{[p]Zureck} W. H. Zureck, Phys. Rev. A 67 (2003) 012320.

\bibitem{[p]Dahlsten} O. Dahlsten, R. Renner, E. Rieper, and V. Vedral. arXiv:0908.0424.

\bibitem{[6]Scully} M. O. Scully, M. S. Zhubairy, G. S. Agarwal, and H.
Walther, Science 299 (2003) 862.

\bibitem{[7]Kim} S. W. Kim and M.-S. Choi, Phys. Rev. Lett. 95 (2005) 226802.

\bibitem{[8]Kim} S. W. Kim and M.-S. Choi, J. Kor. Phys. Soc. 50 (2007) 337.

\bibitem{[8]Raizen} M. G. Raizen, A. M. Dudarev, Q. Niu, and N. J. Fisch,
Phys. Rev. Lett. 94 (2005) 053003.

\bibitem{[9]Marathe} R. Marathe and J. M . R. Parrondo, Phys. Rev. Lett. 104 (2010) 245704.

\bibitem{[9]Serreli} V. Serreli, C.-F. Lee, E. R. Kay, and D. A. Leigh,
Nature 445 (2007) 523.

\bibitem{[10]Thorn} J. J. Thorn, E. A. Schoene, T. Li, and D. A. Steck,
Phys. Rev. Lett. 100 (2008) 240407.

\bibitem{[11]Price} G. N. Price, S. T. Bannerman, K. Viering, E. Narevicius,
and M. G. Raizen, Phys. Rev. Lett. 100 (2008) 093004.

\bibitem{[12]Zurek} W. H. Zurek, in \emph{Frontiers of NonEqiulibrium Statistical
Physics}, edited by G. T. Moore and M.O. Scully, Plenum, New York,
1984.

\bibitem{[13]Lloyd} S. Lloyd, Phys. Rev. A 56 (1997) 3374.

\bibitem{[16]Kim} S. W. Kim, T. Sagawa, S. De Liberato, and M. Ueda, Phys.
Rev. Lett. 106 (2011) 070401.

\bibitem{[14]Bender} C. M. Bender, D. C. Brody, and B. K. Meister, Proc.
R. Soc. A 461 (2005) 733.

\bibitem{[20]Sagawa} T. Sagawa and M. Ueda, Phys. Rev. Lett. 102 (2009) 250602.

\bibitem{[Z]zhang} Y. D. Zhang, \emph{A Grand Dictionry of Physics Problems And
Solutions: Quantum Mechanics}, Science Press, Beijing, 2005.

\bibitem{[11]Kieu} T. D. Kieu, Phys. Rev. Lett. 93 (2004) 140403; Eur.
Phys. J. D 39 (2006) 115.

\bibitem{[12]Esposito} M. Esposito and S. Mukamel, Phys. Rev. E 73 (2006)
046129.

\bibitem{[22]Lev B} L. B. Levitin and T. Toffoli, Theor. Phys. 10 (2011) 1007.

\bibitem{[22]Dong} H. Dong, D. Z. Xu, C. Y. Cai, and C. P. Sun, Phys. Rev. E
83 (2011) 061108.

\bibitem{[p]Ralph} R. Baierlein, \emph{Thermal Physics},
Cambridge University Press, Cambridge, UK, 1999.






\end{thebibliography}
\end{document}